\documentclass{kluwer}    
\newdisplay{guess}{Conjecture}

\begin{document}

\begin{article}
\begin{opening}
\title{Avalanching and Self Organised Criticality, a paradigm for geomagnetic activity?
}
	    \author{Sandra \surname{Chapman$^1$} and Nicholas \surname{Watkins$^2$}}
	    \runningauthor{Chapman and Watkins}
\runningtitle{Avalanching, SOC, and geomagnetic  activity.}
	    \institute{$^1$Univ. of Warwick, Coventry, CV4 7AL, UK\\ 
	    $^2$British Antarctic Survey (NERC), Cambridge, CB3 0ET, UK}
	    \date{}

	    \begin{abstract}

The characterization of global energy storage and release in the
coupled solar wind-magnetosphere system remains one of the fundamental
problems of space physics.
Recently, it has been realised that a  new paradigm in physics, 
that of Self Organised Criticality (SOC) may encapsulate  the mixing
and merging of flux on many scales in the magnetotail prompting bursty energy release
and reconfiguration. SOC is consistent with qualitative measures
such as power law power spectra and  bursty bulk flows and with more quantitative
tests such as power law burst distributions in auroral indices and auroral optical
activity. Here, we present a careful classification of the broad range
of systems that fall under the general description of ``SOC". We argue that some,
but not all,  of these are consistent with our current understanding
of the magnetosphere. We discuss the observed low dimensionality
 of the dynamic magnetosphere in terms of both SOC model properties, and observables.
Observations of burst statistics are highlighted; we show that these are currently
suggestive but not sufficient to confirm SOC and in particular we find that auroral
indices are not effective at distinguishing the internal dynamics of the magnetosphere from
that of the intermittent solar wind driver. This may also elucidate the paradox
of  predictability and complexity of the coupled solar
wind-magnetosphere system.
	    \end{abstract}

\end{opening}

\section{Introduction}

An important avenue of substorm research has been the study of the
  state space of the coupled solar wind-magnetosphere-ionosphere system
  (see e.g. the review  of Klimas et al., (1996)).
The auroral electrojet index (AE) has been used to
  estimate  the   energy released from the magnetosphere into
the ionosphere.  
  One approach
has been the
search for low dimensional chaos in AE, which motivated
  studies of its fractal dimension and Hurst exponent(s).
These studies indicate  that AE has two self-affine regions with
  a break in the scaling exponent at about 2 hours (Takalo et al., 1993).
   Recent work in this area has
   been concerned with identification of chaos in systems for which
  the driver must also be considered; with provision of a robust suite of
  indicators of low dimensional chaos; and with the relationship of these
  methods to the succesful nonlinear predictive filter techniques (see the
  review of  (Sharma, 1995)).

  However, as  noted by Chang
  (1992,1999), low dimensionality is   a property of a system near
  criticality.   Critical behaviour (such as scale invariant fluctuations on all length  scales) had previously been known in systems at
  phase transitions (e.g. Huang (1987)) and hence for small regions  of parameter space.  The ``sandpile" cellular automaton constructed by
   Bak et al. (1987) (BTW)  
   exhibited this behaviour as a natural consequence  of its time evolution, without parameters having to be adjusted ``by
  hand", which was thus dubbed self-organised criticality (SOC). Such a 
  non-equilibrium system dissipates energy by means of many avalanches
  of all sizes, but returns to its out of equilibrium critical 
  stationary state rather than
  relaxing to a non critical one. Hence an alternative framework for
  explanation of the low dimensionality,   
    the
  existence of localized current disruption and bursty bulk flows (e.g. Lui et al., 1988; Angelopoulos et al, 1996) and power law magnetic field spectra in the
  magnetotail (Hoshino et al., 1994), has been given by postulating the magnetotail to be an
  open, dissipative dynamical system at a (forced or self organised) critical state (Chang, 1992;1999; see also Klimas et al. (2000)). 
  In particular, SOC implies that
  the probability distributions of energy release events within the system
  are power laws. Recent evidence for this has been found in auroral images
  (Lui et al. 2000). Also Consolini (1997;1999) has revealed a power law
  in a measure of burst size in AE.
  However, recent analysis by Freeman et al. (2000a) 
  as we shall see, calls into question the usefulness of indices in this context.

Other indicators such as Dst and particle injection
events in the near earth magnetosphere indicate the outflow of
energy from the system due to large scale
`systemwide' reorganizations. These are typically not self  similar,
rather the intensity and time intervals between one substorm and the next
have a probability distribution with a well defined mean 
(e.g. Borovsky et al., 1993; Smith et al., 1996). 
Consistent with this and the SOC conjecture 
 a simple avalanche model  (Chapman et al., 1998,1999)
has been shown to exhibit global events with a characteristic
mean and internal events showing two  scale free regions.

We therefore have a mixture of
requirements for any attempt to model the system, part observational and
part theoretical.
The concepts of self organisation and criticality will first be discussed
 in section 2. We will then attempt  to establish whether this idealized
SOC state is in fact needed to account for the observed
burstiness, self similarity and low dimensionality associated with
magnetospheric dynamics. In section 3 we contrast sandpile models
with other  
models which display intermittency such as Coupled Map Lattices
(CML) which may be understood in terms of dissipative chaos.
 Finally in section 4, we discuss observables that may distinguish  
these  approaches.

\section{Self Organization and Criticality}

One difficulty in applying  SOC is its broad
definition  in the literature.
We  identify three main classes relevant
to magnetospheric physics. One is the original mechanism suggested by Bak et al. (1987) 
which  we will call  ``SOC".
 The second   was introduced by
  Chang (1992) and   referred to  as (Forced) ``F/SOC''. The third is a phenomenological definition
based on observation of some or all of a set of possible 
diagnostics of SOC such as  bursty time series, ``1/f'' power spectra and avalanche
distributions. We shall use the term ``SOC-like'' for these. 
These distinctions  are useful since they refer to different sets of
phenomenology and underlying
theoretical structure.

\subsection{Self Organized Criticality (SOC)}

Noise with very long correlation times, the so-called  ``1/f" or ``flicker'' 
noise is ubiquitous in nature. It is distinguished by a power-law, i.e.
 scale-free, power spectral density of the form $f^{-\beta}$, with $\beta$  
 typically  between $0.8$ and $1.4$. Its spatial counterpart is fractal structure. 
 The apparent ubiquity of these phenomena led BTW to propose  the  SOC mechanism
 as their common origin.   BTW   showed that a discrete cellular automaton simulation (``sandpile model") of a spatially extended, 
slowly driven, many degree of freedom system  in more  than one dimension could
 exhibit both a scale-free spatial response to perturbation and  bursty time evolution.
 They demonstrated the scale-free property in the spatial response by 
exhibiting power-law  ``avalanche'' distributions of size, duration and 
spatial extent of toppling events in the model.    The mechanism that  BTW 
proposed to explain this  behaviour  was an 
underlying fixed point in the dynamics (``criticality"), which was attractive 
(``self-organised").  
Support for the presence of  critical behaviour in the original
BTW model was given mainly by their demonstration
 of finite-size scaling in the avalanche distributions for 
 different lengths of the model system, as this was a property
  unique to critical systems.   

Demonstration (particularly by renormalisation group (RG) methods) that sandpile
models can indeed exhibit  attractive fixed points  has thus been an important thread in SOC research for the past decade (see
(Jensen, 1998) for a brief review).  Such fixed points have been observed
in the conservative BTW model and others,  hence SOC has been shown to
exist as a mechanism. A parallel thread has
been the experimental search for SOC in real sandpiles, ricepiles, and other rather less controllable systems such as seismic
faults, solar reconnection events and astrophysical accretion disks (see the reviews of Jensen, 1998; Turcotte, 1999
and Dendy and Helander, 1997). The only  signature of criticality  per se 
  so far tested for in data is finite-size scaling,
which requires the system size to be changed,
so only the Oslo ricepile (Frette et al., 1996) 
among natural systems can   be said to demonstrate SOC behaviour in this sense. The relevance of 
criticality in particular for magnetospheric dynamics is that such a critical
 system effectively becomes low dimensional, as noted by  Chang (1992).
In addition it may have implications for the presence or absence 
of time correlation between successive energy release events, as conjectured
by Watkins et al. (2000), Consolini (1999) and 
Freeman et al., (2000b) following  Boffetta et al., (1999).

\subsection{F/SOC}

Although BTW hypothesised a mechanism (SOC) to explain the
avalanche phenomenology they observed, it remains  possible that many natural
 or model systems may share all or part of the avalanche phenomenology  
  without necessarily being  self-organised.  
One difference from SOC comes with some model systems
including  the ``Forest Fire'' models  which are controlled by repulsive rather than attractive
  fixed points, and so have to be  tuned 
  to exhibit scaling, rather than being attracted there from arbitrary initial conditions in the control parameters. 
Chang has called this ``forced criticality",
and it differs from SOC as it is critical without self-organisation.
It has relevance to the problem of magnetospheric dynamics,
as argued by Chang, (1992;1999) and also Consolini and de Michelis, (2000) since the magnetospheric system may be driven to a critical or near-critical state as
a result of the continuous loading process that it undergoes.
Indeed, the simple model used by Chapman et al (1998)  has been shown to possess 
a repulsive fixed point by RG (Tam et al, 2000).

\subsection{SOC-like}

The description ``SOC-like'' may be usefully applied
 for systems which have not yet been analytically treated
and for which the classic tests of criticality such as finite size scaling 
either cannot be or have not yet been performed. A good example 
of the latter are the thresholded diffusion equations which have received extensive study 
(see section 5.4 of  (Jensen, 1998))  as continuous space and time differential models for 
avalanche phenomenology. An important example  is that of  Lu (1995)
 who
demonstrated that the differenced diffusion equation for an arbitary one
dimensional field $\phi$
\begin{equation}
\frac{\partial \phi(x,t)}{\partial t} =
\frac{\partial}{\partial x} [ D(x,t) \frac{\partial \phi}{\partial x} ] + S(x,t) 
\end{equation}
with a noise-like driver $S(x,t)$ and a particular choice
of nonlinear diffusion coefficient 
$D(x,t)$ generated avalanching with power law statistics. 
 One can then (Vassiliadis et al., 1998)  cast the MHD equation with variable $\eta$
\begin{equation}
\frac{\partial {\bf B}}{\partial t} = - \nabla \wedge (\eta {\bf J})
+ \nabla\wedge (\bf{v} \wedge  {\bf B})
\end{equation}
in terms of diffusion dominated MHD, by replacing the advection term by
a  source term without explicit velocity dependence
giving 
\begin{equation}
\frac{\partial {\bf B}}{\partial t} = - \nabla \wedge (\eta {\bf J})
+ S({\bf x}, t),
\end{equation}
 then using $\mu_0 J = \nabla \wedge {\bf B}$ and solving for ${\bf B}$ using a
discrete space-time but continuous field scheme.
This method then retains the phenomenology of resistive MHD provided that diffusion 
initiated mixing and merging dominates convection - it has yet
to be demonstrated that this is an appropriate approximation for the 
magnetosphere. The above model
has been further developed by Klimas and co-workers (see Takalo et al., 1999;
Klimas et al., 2000 and references therein). 
 However, equation (3) in one dimension corresponds essentially to
 Burgers' equation with a noise drive, depending upon the
 form of $\eta$. Care hence needs to be taken with
 solutions of this system to distinguish Burgers turbulence-type solutions from those that could only be SOC- like (Krommes, 2000).
 Solutions of the
 diffusion equation are naturally long-range in space and time without invoking criticality. 
  However the avalanche 
 distributions that are also observed may indeed indicate critical or near-critical  behaviour
and we may thus call 
  them SOC-like pending analytic study.
Such questions have recently become more relevant since
 several authors (in particular Boffetta et al., (1999)) have noticed the similarity of avalanche distributions  in SOC systems to the distributions of amplitudes and lifetimes in shell models of turbulence (see section 3.1 below).

Finally, it should be noted that the low dimensionality of the dynamic
magnetosphere supports phenomenology that may be  ``SOC-like" rather 
than SOC.
Singular spectrum analysis of previously identified substorm events
in auroral indices (see for example Sitnov et al., 2000)
reveals a state space that is low dimensional
and may be described reasonably well by a cusp catastrophe, as first
suggested by (Lewis, 1991). 
In this context it is critical to understand to what extent measures of
the system  dynamics such as
auroral indices  also measure the solar wind drive
directly and  hence to quantify their appropriateness  for 
such studies; we shall see in section 4 this remains an open question.

\section{Models}
\subsection{Numerical Models for Avalanching and Intermittency}

 A generic approach is to obtain a reduced description of
 a given system that can be implemented on a lattice.
 Briefly,  lattice based numerical models may be classified (Bohr et al., 1998)as i) reduced
 forms of the governing PDE as in the case of turbulent shell models
 that are obtained by approximate truncation of  the Navier Stokes equations,
 ii) Coupled  Map Lattices that spatially couple systems with prescribed nonlinear time
 evolution to give
 simple models of spatiotemporal chaos, and iii)
 avalanche models that combine simple redistribution rules (spatial coupling)
 with thresholded time evolution. 

Many variants of turbulence shell models exist and we refer the reader to
(Bohr et al., 1998) for a comprehensive discussion. Briefly,
these consist of difference equations in $k$ space that
couple $k$ modes locally thus allowing energy to cascade in $k$
which can generate power law power spectra with
Kolmogorov scaling. It is relevant to note here
that shell models such as that of Gledzer, Ohkitani and Yamada (the GOY
model, see Bohr et al., 1998) generate intermittency and power law power spectra and yet must be distinguished from other
numerical systems with bursty evolution.  Power-law distributions in lifetime and burst size
have been observed  in a shell model by  Boffetta et al., (1999).
The  observation by Einaudi and Velli, (1999) of an avalanche 
 distribution arising from   MHD plasma simulations of anisotropic turbulence  
 suggests that 
 the avalanche distribution may be largely insensitive to the 
 underlying physics of the turbulent system.  

Coupled Map Lattices (CML), on the other hand, model spatiotemporal
chaos in configuration space. Although many variants have been explored
(see Kaneko, 1993 for a review) the essential philosophy is to 
decompose the processes underlying the phenomena of interest into
independent components (e.g. convection, diffusion) each of which
may be nonlinear, and then to reduce each of these to simple 
parallel dynamics on a lattice. A typical CML may then comprise
a local nonlinear map (Logistic map, say) evolving each variable in
time at each lattice point on the grid, plus coupling between the points
(discretized, and therefore truncated, diffusion, say) which is linear.
Spatiotemporal chaos in this sense then implies deterministic chaos that is high dimensional;
because the number of dimensions increases as we increase the number of coupled
nonlinear nodes on the lattice. The deterministic nature of
the local evolution means that Lyapunov exponents are well defined for
the coupled system in contrast to most ``sandpile" algorithms. In addition certain universality classes, including that
of diffusive coupling, can be constructed to be tractable via RG. 
The generation of intermittency via this ``classic" CML should therefore
be distinguished from other grid-based real variable sandpile models
which are also sometimes
dubbed ``CML'' but which, due  either to the introduction
of randomness, or thresholds for diffusion, or both, do not have
well defined Lyapunov exponents and cannot be treated as high dimensional
chaotic systems in this sense. This definition of the ``classic" CML 
can also be seen to be related to, but distinct from, low dimensional chaos
which can also generate intermittency; both show intermittency for particular
values of control parameters in their underlying nonlinear maps. The low dimensional
system may in principle be distinguished via phase space reconstruction, as has
been attempted for the magnetosphere (Sharma, 1995).

Several avalanche models have been studied in the 
magnetospheric context, and we briefly recap here.
In addition to the Lu-type models
discussed in section 2.3.
 Consolini (1997, 1999), studied the 1D Hwa-Kardar ``running'' algorithm which 
 gives a broken power
law power spectrum similar to that of $AE$.
A 2D modification of the BTW model was used by  Uritsky and Pudovkin, (1998), 
in which the thresholding rule was ``driven'' using the $AE$ index. Chapman et al., (1998,1999)
 have studied a 1D model
with a continuous variable. In distinction to the cellular automata this
may be called a ``discrete space time'' model. This model had nonlocal redistribution rules and thus manifested nontrivial distributions even in one dimension, (see also Uritsky and Semenov, 1998).
More recently Consolini and de Michelis (2000) have studied a 2D modified Forest Fire model driven by a 1D coupled map lattice.

\subsection{Models for Self organised Low Dimensionality}

One resolution of the observational evidence for low dimensionality in the
dynamic magnetosphere, and bursty evolution that is also robust (and hence
less easy to explain in terms of deterministic chaos) is to note that systems
at criticality can also be low dimensional. This key point was raised by Chang
(1992), and  the implication is the following.  Avalanche (sandpile)
models, forest fire models and so forth have robust emergent phenomenology that
yields bursty time evolution with power law burst statistics as required but
these systems are by construction high dimensional, in the same sense
as CML. 
If in addition these systems exhibit fixed points, then close to the fixed points,
that is, close to criticality, the behavior is low dimensional.
For avalanche models to describe the low dimensional magnetosphere they must
have fixed points. This places a rather strong restriction on the
models which may, for example, exclude those proposed to more closely resemble diffusion
dominated MHD.  

In this context it is intriguing to note that, just as classical CML,
avalanche models can be modified to exhibit low dimensional behaviour.
We give an example here by modifying the sandpile model used by
(Chapman et al., 1998). Briefly, this model is similar to others in that sand is redistributed
when a critical gradient is exceeded locally. Unlike other
models however, redistribution occurs
across all sites within an ongoing avalanche by construction; 
the sand is redistributed conservatively such that within the avalanche the gradients
are all set to zero. We can consider a generalization of this model by
introducing a ``fluidisation parameter" that is, by flattening back the sand behind
the leading edge of an ongoing avalanche for a fixed distance $L_f$. This
has been extensively investigated by Chapman (2000) and corresponds to
moving the system away from the repulsive fixed point. Here we present 
a summary of the results which are germane to the discussion. The behaviour of the system
with $L_f$ essentially has two regimes, when $L_f$ is of order the system size, the 
behaviour is just that of the original $L_f=N$ (system length) sandpile model
considered in Chapman et al. (1998;1999), that is, evolution is bursty
and burst statistics are power law. However, if we reduce
$L_f$ to much less than a quarter of the system size, although energy can still
only be released by avalanches, the evolution becomes quasiregular, with
a distinct loading-unloading cycle and statistics that are power law only
over a restricted range. The essential point here is that an originally
high dimensional sandpile model can, for appropriate parameters, exhibit low
dimensional dynamics. We show in Figure 1 the time series $E(t)$ for these two cases.
For $L_f=2000$, $N=4096$, the system evolution is  ``intermittent" whereas
for $L_f=4, N=4096$ it is quasiregular, and for this latter case we show in Figure 2 the
limit cycle for the system obtained by embedding. 
The implication is that low dimensionality can either be a signature of a system close
to criticality or of certain classes of avalanching systems that can be
tuned to give either intermittent, or quasiregular, time evolution.

\section{What Can We Measure?}

For models and laboratory experiments the identification of
SOC or SOC-like phenomenology is fairly straightforward.
One would expect evidence in burst size and life time distributions of
power laws; evidence of fractal structures and long range 
correlations; and finite size scaling as one varies the system size.
Unfortunately, dealing with observations rather than carefully
constructed experiments presents some nontrivial problems and we highlight these
next. Of principal concern in the  magnetosphere is the variability of the driver and
the extent to which any given observable yields the output of the system, the 
system's internal dynamics, or a mix of these with the driver superimposed. 
The problem of deciding when an observation is characterising the driver, or 
the system's response to it, is problematic since the driver in this case, that is, the solar wind,
is known to be turbulent and exhibit intermittency. One approach in terms of
 modelling (Consolini and de Michelis, 2000) is to couple the avalanche model 
to an intermittent drive; this may mimic the overall coupled 
driver-system behaviour but does not easily allow one to unravel the complex dynamics of the drive from that of the system.

Auroral indices have received considerable attention as a means for testing
the predictions of models showing complexity. These indices have the advantage that they
provide long time series; essential to test the hypothesis of scale free (power law)
behaviour. The broken power law form of the power spectrum of AE (Tsurutani
et al, 1990) and later, its burst lifetime and 
size distribution (Consolini,1997;1999) have been taken to be strong indicators of complexity and 
SOC in the magnetosphere's dynamic evolution. Comparisons between the indices
and the solar wind drive, with an emphasis on predicting magnetospheric
activity have however implied that the dynamics is to some extent predictable
(Sharma,1995). Care must be taken to clearly define  what actually constitutes sucesful
prediction of some observed activity;  this also relies on some clear definition
of substorm onset and differentiation between substorms and other sporadic
events (BBFs, pseudobreakups). In terms of SOC phenomenology the question is more clearly  adressed
 if we compare burst distributions from the magnetosphere and the solar wind.

 Consolini (1997,1999) and Takalo et al. (1999; see
also references therein) have defined a measure of energy bursts 
i.e. integrated dissipation rate with respect to an arbitrary constant threshold level and applied it to the auroral electrojet index $AE$.  They
 found power laws in both size and burst duration, but Consolini (1999), has
 since shown the presence of a small ``bump" with  characteristic 
 values of burst magnitude $e$ and duration $T$. Freeman et al., (2000a) showed that the burst lifetime
 distributions $P(T)$ for  $AU$ and $AL$ between January 1978 and June 1988
 share this form. 
In addition observations from the WIND satellite's
 particle and magnetic field instruments between January 1995 and
December 1998  were used  to show that 
the power law component of $P(T)$ but not the ``bump"
is  also present  in two measures (Akasofu's $\varepsilon$  and $v B_s$)
which estimate the energy delivered by the solar wind to 
the Earth's magnetopause region. This close correspondence 
is illustrated in Figure 3  from
Freeman et al. (2000a) comparing  curves for $AU$ and $\varepsilon$. Freeman et al., (2000a)
presented arguments that the ``bump" corresponds to the effect of the  DP1 substorm ``unloading''
current system.

Evidence has since been  given (Freeman et al., 2000b) suggesting  that the scaling region
common to $AU/AL$ and $\varepsilon$ is in fact present in the energy flow within the
solar wind itself, as measured by $P(e)$ and $P(T)$ for the Poynting vector. 
In addition (Freeman et al., 2000b) a power law inter-burst interval distribution
has been exhibited for the Poynting vector which
in turn has implications for the predictability 
of fluctuations in the DP2 ionospheric current system, if later work 
confirms that the bursts in $AU/AL$ are
causally related to those in $\varepsilon$ and $v B_s$. 
We note that Consolini (1999) has shown
AE to also have a power-law rather than exponential 
inter-burst interval distribution.

Another measure of magnetospheric output is auroral optical activity. POLAR UVI
allows almost the entire auroral oval to be imaged. These images show
substorms plus smaller events which may be related to BBFs and other
bursty reconfigurations within the magnetotail. One can construct a burst distribution
for these and such a study was performed by Lui et al. (2000). 
The key results are shown in Figure 4. Here ``blobs" of brightness in the auroral
oval have been identified and, after suitable background subtraction, both
the intensity summed over a blob, and its area, can be obtained. When the
probability distributions of integrated intensity and size were plotted Lui et al (2000)
found a power law slope plus a ``bump" at large values corresponding to substorm
breakups. Looking at quiet times where no substorms occur they found
the power law with unchanged index only. This appears to be strong evidence that the non-substorm
(the internal) events are power law whereas the substorm (systemwide) events are
not, consistent with simple avalanche models such as (Chapman et al., 1998). 

These two observational results may not be in conflict if we consider that both the
drive and the internal magnetospheric dynamics are SOC-like, because it has also been shown
(Chapman et al., 1999) that avalanche models may tend to absorb information about detailed fluctuations
in the drive. However, the striking coincidence between the burst distributions
of the indices and of the solar wind found by Freeman et al. (2000a),
for all except the largest events is suggestive that auroral indices are, for
the small scale (non-systemwide) events, strongly influenced directly by the drive itself, and are therefore
i) easy to predict and ii) not a good method for uniquely testing the SOC hypothesis  for magnetospheric dynamics. 

\section{Conclusions}

Here we have elucidated the different classes of SOC description, and related
descriptions of turbulent and other high dimensional systems
suc as CML. This has suggested
that 
 that most, if not all, of the current evidence for
SOC,F/SOC or SOC-like behaviour in the magnetosphere, both quantitative
 (avalanche distributions, power law spectra, low dimensionality and self-affinity), and more qualitative (burstiness and inherent multiscale nature)
could be obtained from either 
shell models for turbulence, {\em bona 
fide} coupled map lattices or avalanche models.
A particular ambiguity is that the low dimensionality which Chang (1992)
pointed out was a feature of forced  criticality, can also
arise from low dimensional chaos, or from a `tuned' high dimensional chaotic system under certain conditions.
These all however, embody different aspects of the underlying physics and
therefore future work needs to be directed specifically at distinguishing
them; in particular multispacecraft missions such as CLUSTER II
will allow unambiguous turbulence measures, as well as avalanche
statistics to be determined in situ.

 A  difficulty is that to test for power law dependence, long timeseries
 or large sets of observations are needed. In principle, geomagnetic indices
 are suitable in this respect, but recent work
suggests that the power law part of the avalanche distributions
shown by magnetospheric indices may in fact be a directly driven
manifestation of solar wind control of the DP2 current.
This also prompts a reexamination of other uses of indices
to unravel the relationship between the solar wind driver and
the response of the dynamic magnetosphere.

{\em Note added in proof} \\
It is important to distinguish the meaning of an ``attractive fixed
point" in the {\em time evolution} (phase space or parameter space) of a
 system from that in the {\em flow of the  real space renormalisation group}
 transformation applied to a  critical or near critical system. Insofar as it 
 evolves to criticality from arbitrary initial conditions in parameter space, 
 an \underline{attractive} fixed point in  the former sense has been used to 
 explain the ``self-organization" of SOC systems (e.g. Jensen, 1998). A 
 \underline{repulsive}  fixed point under a real-space RG  transformation, 
 however, is generally found for critical systems and  is a diagnostic for 
 phase transitions (e.g. page 79 and chapter 11 of Sornette, 2000).

{\bf Acknowledgements}

SCC acknowledges a PPARC lecturer fellowship. The authors acknowledge valuable discussions with 
Tom Chang, Mervyn Freeman and George Rowlands. NWW is grateful to MIT for hospitality during
the writing of this article.



\newpage
\newpage
\begin{figure}
\caption{Timeseries of energy released during avalanches with flattening
back length a)$L_f=4$ and b)$L_f=2000$}
\end{figure}

\begin{figure}
\caption{Phase space reconstruction
for the timeseries shown in part in figure 1(a)}
\end{figure}

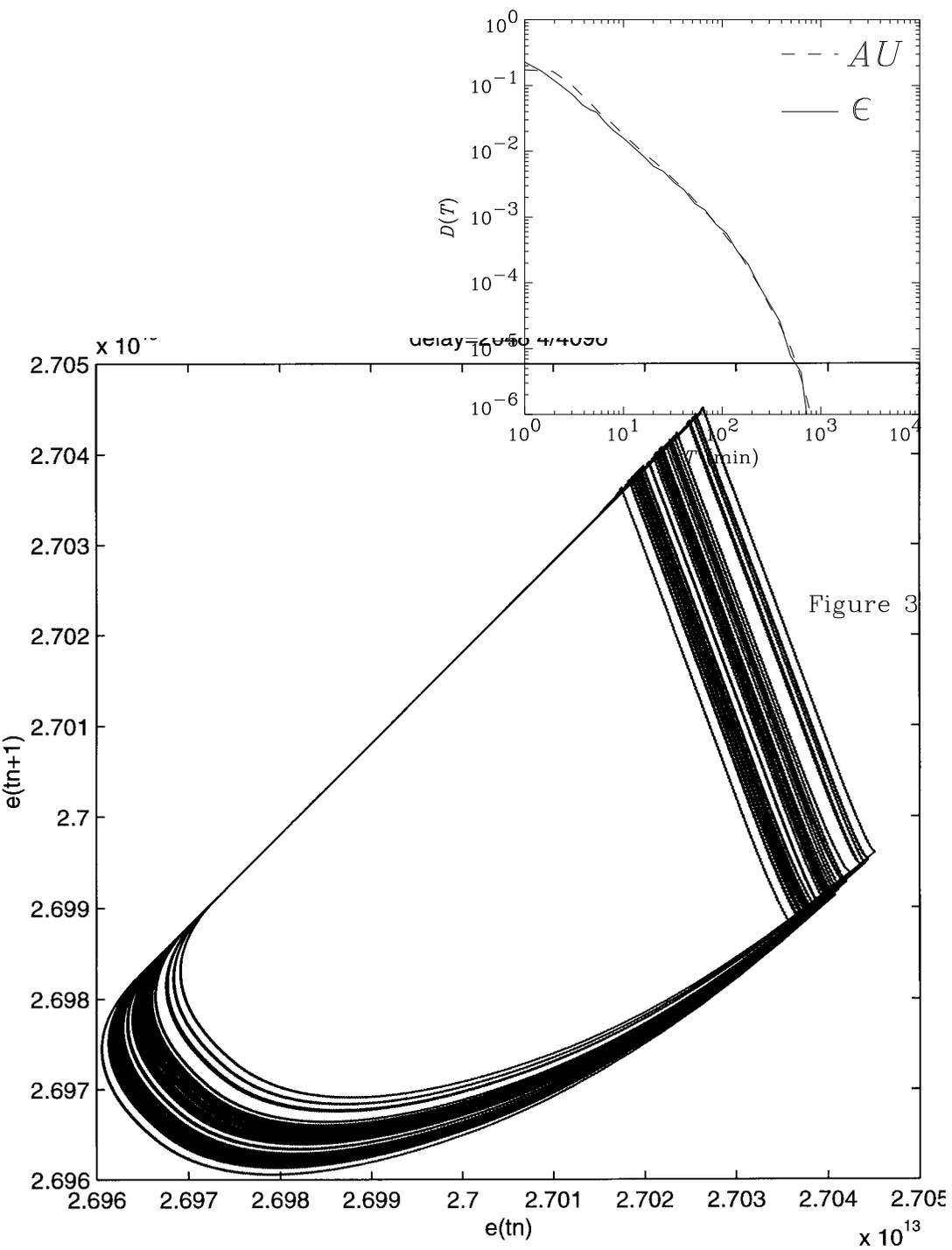
\begin{figure} 
\caption{A direct comparison of the burst lifetime PDFs of $AU$ 
(for January 1978- June 1988) and $\varepsilon$ (calculated from WIND SWE and  MFI data 
 for 1984-1987). Thresholds have been chosen to give similar exponential cutoffs
 to lifetimes. From Freeman et al. (2000a)}
\end{figure}

\begin{figure}
\caption{Probability distributions of the size, and energy dissipated, in
auroral 'blobs' during quiet and active times, from Lui et al. (2000)}
\end{figure}

\end{article}

\begin{thebibliography}{}

 
\bibitem{}
 Angelopoulos V., Coroniti, F. V.,  Kennel, C. F., Kivelson, M. G.,
 Walker, R. J., Russell, C. T., McPherron, R. L., Sanchez, E.,  Meng, C. I.,
 Baumjohann, W., Reeves, G. D.,  Belian, R. D.,  Sato, N., Friis-Christensen, E., 
  Sutcliffe, P. R., Yumoto, K.,  and Harris, T.: 1996,
   {\em J. Geophys. Res.} {\bf 101}, 4967.

\bibitem{} 
Bak, P.,  Tang, C., and Weisenfeld, K.:1987, `Self--organized criticality: An
explanation of 1/f noise', {\it Phys. Rev. Lett.} {\bf 50}, 381.


\bibitem{}
Bohr, T., Jensen, M., Paladin, G., and Vulpiani, A.: 1998, {\em
Dynamical Systems Approach to Turbulence}, Cambridge University Press,
p. 350.

\bibitem{}
Boffetta, G., 1999, Carbone, V.,
Giuliani, P., Veltri, P., Vulpiani, A.: 1999,
`Power laws in solar flares: self-organized criticality or turbulence ?',
{\em Phys. Rev. Lett.} {\bf 83}, 4662. 


\bibitem{}
Borovsky, J. E., Nemzek, R. J., and Belian, R. D.: 1993,
  `The occurence rate of magnetospheric-substorm
onsets: Random and periodic substorms', {\em J. of Geophys. Res.} {\bf 98}, 3807.

\bibitem{} 
Chang, T. S.: 1992, `Low dimensional behaviour and symmetry breaking of
stochastic systems near criticality - can these effects
be observed in space and in the laboratory ?', {\it IEEE Trans. Plasma Sci.}, {\bf 20}, 691.

\bibitem{}
Chang, T. S.: 1999, `Self-organized criticality, multi-fractal spectra,
 sporadic localized reconnections and 
intermittent turbulence in the magnetotail', {\it Phys.  Plasmas} {\bf 6}, 
4137.

\bibitem{}
Chapman, S. C.: 2000, `A deterministic avalanche model with limit cycle
exhibiting period doubling, intermittency and self
similarity', {\em Phys. Rev. E}, {\bf 62}, 105.

\bibitem{} Chapman, S. C., Watkins, N. W., Dendy, R. O.,
Helander, P., and Rowlands, G.: 1998. `A simple avalanche model as an analogue for magnetospheric activity', {\em Geophys. Res. Lett.} {\bf 25}, 2397. 

\bibitem{}  Chapman, S. C., Dendy, R. O., Rowlands, G.:1999. `A sandpile model with dual scaling regimes for laboratory, space and astrophysical
plasmas', {\em Phys. Plasmas} {\bf 6}, 4169.

\bibitem{}  Christensen, K., Olami, Z.,  and Bak, P.: 1992. `Deterministic 1/f noise in nonconservative models of self-organized
criticality', {\em Phys. Rev. Lett.} {\bf 68}, 2417.

\bibitem{} 
 Consolini, G.: 1997, `Sandpile cellular automata and magnetospheric
 dynamics', in    S. Aiello, N. Iucci, G. Sironi, A. Treves and U. Villante (eds.),  
 {\it Proc. vol. 58, ``Cosmic Physics in the  Year 2000"}, SIF, Bologna, Italy.

\bibitem{}
Consolini, G.: 1999, `Avalanches, scaling and criticality in magnetospheric dynamics', 
{\em Phys. Rev. Lett.}, submitted.

\bibitem{}
Consolini, G., and de Michelis, P.: 2000, `A revised forest-fire automaton for
the nonlinear dynamics of the Earth's magnetotail', {\em J. Atmos. Sol-Terr. Phys.}, in press.
     
\bibitem{}
Dendy, R. O., and Helander, P.: 1997, `Sandpiles, silos and tokamak
phenomenology: a brief review', {\em Plasma Phys. Controlled Fusion},
{\bf 39}, 1947.

\bibitem{}
Einaudi, G., and Velli, M.: 1999, `The distribution of flares, statistics 
of magnetohydrodynamic turbulence and coronal heating', {\em Phys.
Plasmas} {\bf 6}, 4146.

\bibitem{}
Freeman, M. P., Watkins, N. W., and Riley, D. J.: 2000a. `Evidence for a solar wind origin of the power law burst lifetime distribution of the AE indices',
{\em Geophys. Res. Lett.}, {\bf 27}, 1087.

\bibitem{}
Freeman, M. P., Watkins, N. W., and Riley, D. J.: 2000b. `An SOC-like avalanche distribution 
observed in an MHD turbulent cascade in the solar wind', {\em Phys. Rev. E.}, in press.

\bibitem{}
Frette, V., Christensen, K., Malthe-Sorenssen, A., Feder, J.,
Jossang, T., and Meakin, P.: 1996, `Avalanche dynamics in a pile of rice',
 {\em Nature} {\bf 379}, 49.

\bibitem{}
Hoshino M., Nishida, A., Yamamoto, T., Kokubun, S.: 1994,
`Turbulent magnetic field in the distant magnetotail: bottom-up process of plasmoid formation', {\em Geophys. Res. Lett.}  {\bf 21}, 2935.


\bibitem{}
Huang, K.: 1987., {\it Statistical Mechanics}, Second Ed., Wiley, New York.
  

\bibitem{}
 Jensen, H.  J.: 1998. {\em Self-Organised Criticality:  Emergent Complex Behaviour in Physical and Biological Systems}, Cambridge University Press, Cambridge, p.
153.

\bibitem{}
Kaneko, K.: 1993, {\em Theory and applications of coupled map lattices},
Wiley, New York.

   \bibitem{}
 Klimas, A. J., Vassiliadis, D., Baker, D. N., and Roberts, 
D. A.:1996, 'The organised nonlinear dynamics of the magnetosphere',
   {\em J. Geophys. Res.} {\bf 101}, 13089.

\bibitem{}
Klimas, A. J.,  Valdivia, J. A.,  Vassiliadis, D.,  Baker, D. N., Hesse, M.,
 and Takalo., J.,: 2000, `The role of self-organized criticality in the substorm phenomenon and its relation to localized reconnection in the magnetospheric plasma sheet', {\em J. Geophys. Res.}, submitted.


\bibitem{}
Krommes, J. A: 2000, `Renormalized dissipation in the nonconservatively
forced Burgers equation', {\em Phys. Plasmas} {\bf 7}, 1064.

\bibitem{} 
Lewis, Z. V.: 1991, `On the apparent randomness of substorm onsets',
{\em Geophys. Res. Lett.} {\bf 18}, 1627.

\bibitem{}
Lu, E. T.: 1995, `Avalanches in Continuum Dissipative Systems', {\em
Phys. Rev. Lett.} {\bf 74}, 2511. 

\bibitem{} 
Lui, A. T. Y., Lopez, R. E., Krimigis, S. M., McEntire, R. W., 
Zanetti, L. J., Potemra, T. A.: 1988, 
`A case study of magnetotail current sheet disruption and
diversion', {\em Geophys. Res. Lett.} {\bf 15}, 721.

\bibitem{}
Lui, A. T.Y., Chapman, S. C., Liou, K., Newell, P. T., Meng, C. I., 
Brittnacher, M., and Parks, G. K.: 2000, `Is the dynamic magnetosphere
an avalanching system ?', {\em Geophys. Res. Lett.}, 27, 911.

   \bibitem{}
Sharma, A. S.: 1995, `Assessing the magnetosphere's nonlinear
   behaviour: its dimension is low, its predictability high', {\it Rev. Geophys.}, 
   {\bf 33}, Part 1, Suppl. S., 645.

\bibitem{}
Sitnov, M. I.,  Sharma, A. S.,  Papadopoulos, K.,  Vassiliadis, D.,
 Valdivia, J. A., Klimas, A. J., and Baker, D. N.: 2000.
 `Phase transition-like behavior of the magnetosphere during substorms',
  {\em J. Geophys. Res.}, {\bf 105}, 12955.

\bibitem{}
 Smith, A. J., Freeman, M. P., and Reeves, G. D.: 1996,
   `Postmidnight VLF chorus events, a substorm signature observed at
   the ground near L=4', {\em J. Geophys. Res.} {\bf 101}, 24641.
  
\bibitem{}   

Sornette, D.:2000, {\em Critical Phenomena in Natural Sciences. Chaos, Fractals,
Selforganization and Disorder: Concepts and Tools}, 
Springer-Verlag, Berlin.

   \bibitem{}
 Takalo, J., Timonen, J., and Koskinen, H.: 1993,
 `Correlation dimension   and affinity of {\it ae } data and bicolored noise', 
{\em Geophys. Res. Lett.} {\bf 20}, 1527.

\bibitem{}
 Takalo, J., Timonen, J., Klimas, A., Valdivia, J., and Vassiliadis, D.: 1999,
 `Nonlinear energy dissipation in a cellular automaton magnetotail
field model', {\em Geophys. Res. Lett.} {\bf 26}, 1813.

\bibitem{}
Tam, W. Y., Chang, T. S., Chapman, S. C., and Watkins, N. W.:2000,
`Analytical Determination of Power Law Index for the Chapman et al. Sandpile
(FSOC) Analog for Magnetospheric Activity-Renormalization Group Analysis', 
{\em Geophys. Res. Lett.} {\bf 27}, 1367.

\bibitem{}
 Tsurutani, B., Sugiura, M., Iyemori, T., Goldstein, B. E.,
   Gonzalez, W. D.,  Akasofu, S.-I., and  Smith, E. J.: 1990, `The nonlinear 
   response of AE to the IMF  $B_s$: A spectral break at 5 hours',
     {\em Geophys. Res. Lett.} {\bf 17}, 279.

\bibitem{}
Turcotte, D. L.: 1999, `Self-organized criticality', 
{\em Rep. Prog. Phys.} {\bf 62}, 1.

\bibitem{}
Uritsky, V. M., and V. S. Semenov, A sandpile model for global statistics of
reconnection events in the magnetotail, Proc. international workshop on
``The solar wind-magnetosphere system 3'', 23-25 September, 1998, Graz,
Austria. 

\bibitem{}
Uritsky, V. M., and Pudovkin, M.: 1998, `Low frequency 1/f-like fluctuations
of the AE-index as a possible manifestation of self-organised criticality in the magnetosphere', {\em Ann. Geophys}, {\bf 16}, 1580.

\bibitem{}
Vassiliadis, D., Anastasiadis, A., Georgioulis, M., Vlahos, L.: 1998, 
`Derivation of solar flare cellular automata models from a subset of the magnetohydrodynamic
 equations', {\em Astrophys. J.} {\bf 509}, L53.

\bibitem{}
Watkins N. W.,  Freeman, M. P.,  Chapman, S. C.,
 and Dendy, R. O.: 2000, `Testing the SOC hypothesis for the magnetosphere',
{\em J. Atmos. Sol-Terr. Phys.}, in press.



 

  
\end{thebibliography}
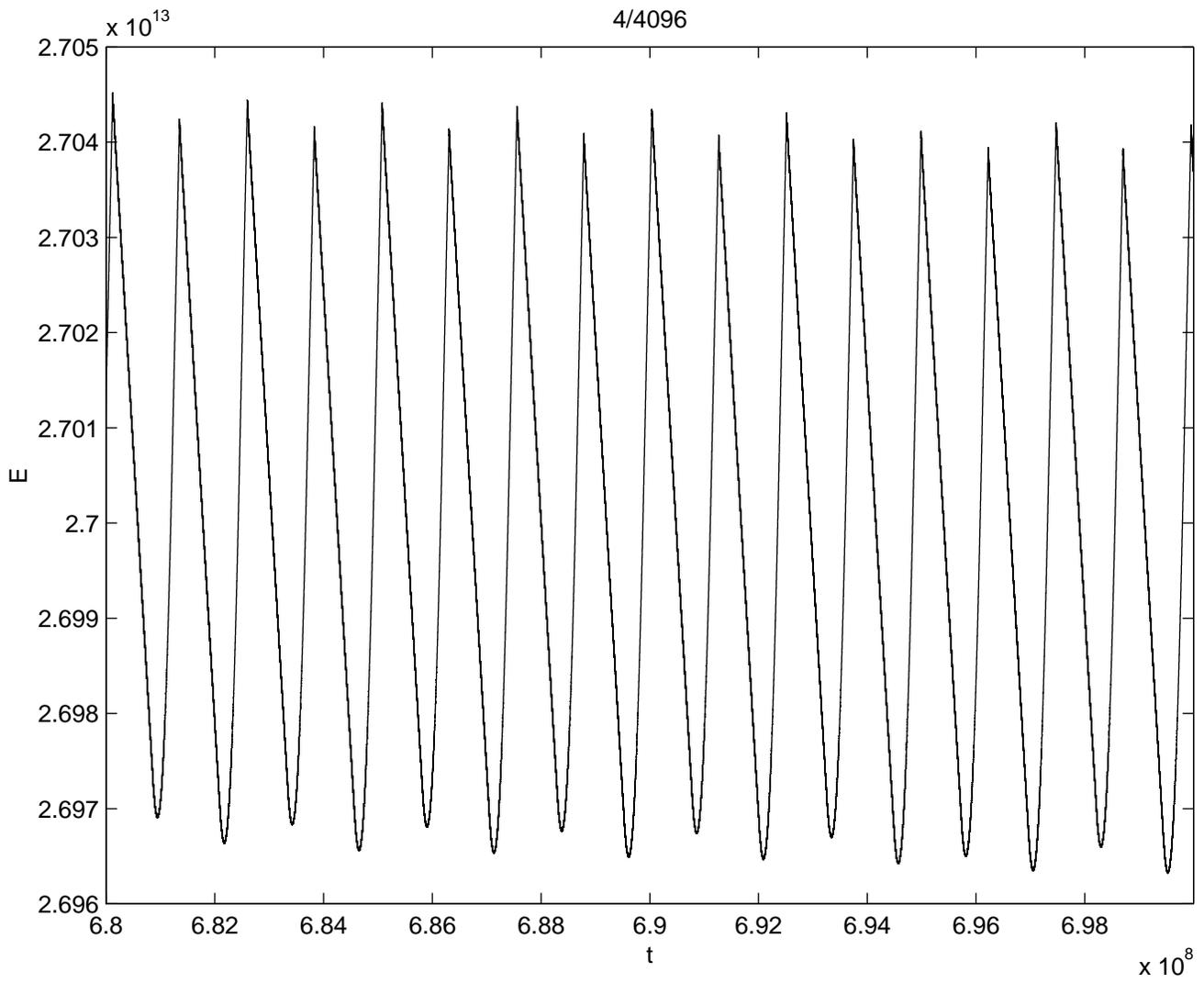
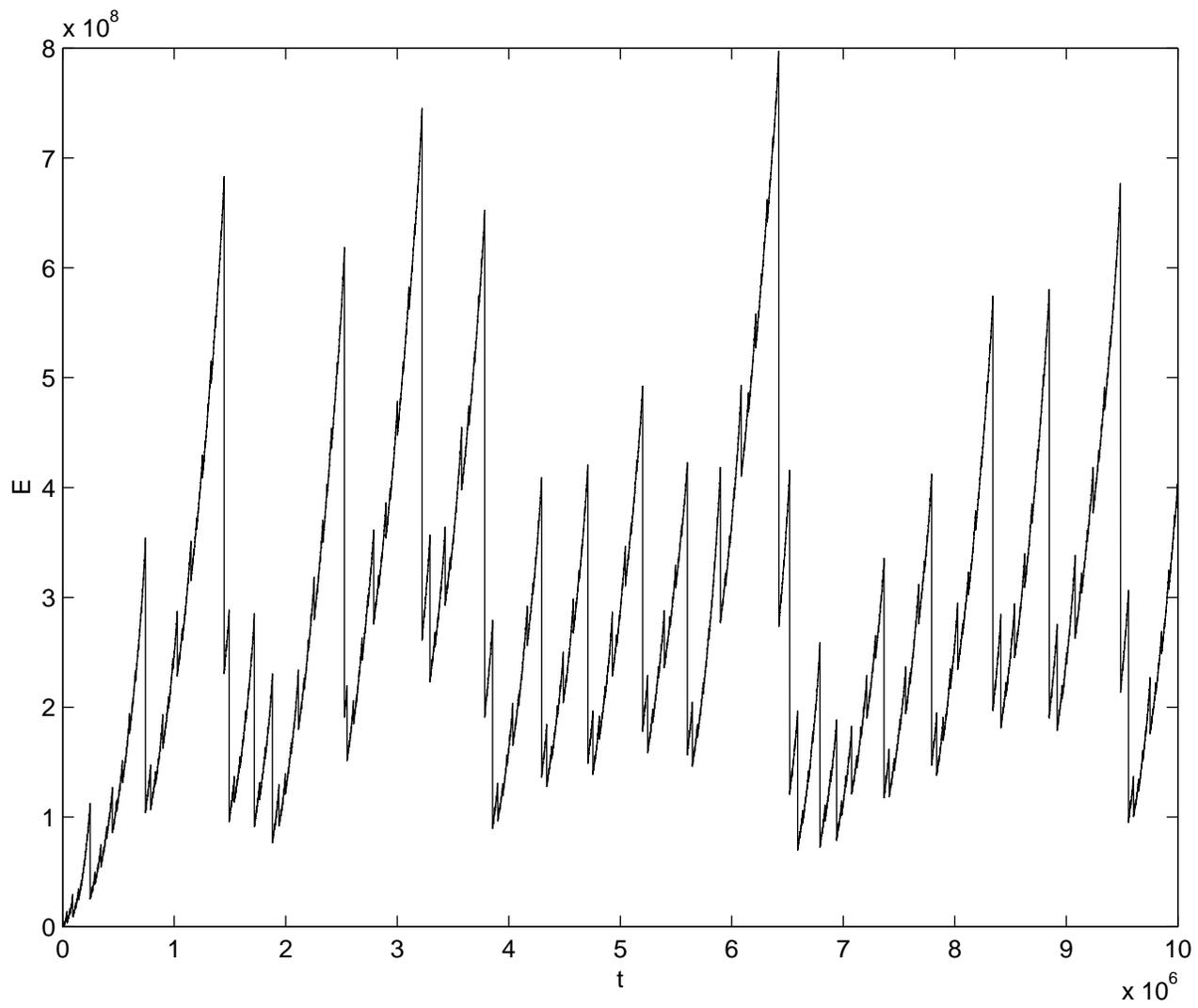
\end{document}